\documentclass[twocolumn,prl]{revtex4}
\usepackage{graphicx}
\newcommand{\be}{\begin{equation}}
\newcommand{\ee}{\end{equation}}
\newcommand{\bea}{\begin{eqnarray}}
\newcommand{\eea}{\end{eqnarray}}

\newcommand{\la}{\langle}
\newcommand{\ra}{\rangle}

\newcommand{\lp}{\left(}
\newcommand{\rp}{\right)}

\renewcommand{\phi}{\varphi}
\renewcommand{\epsilon}{\varepsilon}

\begin{document}

\title{Quantized Transport in Graphene p-n Junctions in Magnetic Field}
\author{
D. A. Abanin and L. S. Levitov
}
\affiliation{
Department of Physics,
 Massachusetts Institute of Technology, 77 Massachusetts Ave,
 Cambridge, MA 02139
}


\begin{abstract}

Recent experimental work on locally gated graphene layers resulting in 
p-n junctions have revealed quantum 
Hall effect in their transport behavior. 
We explain the observed conductance quantization 
which is fractional in the bipolar regime
and integer in the unipolar regime
in terms of quantum Hall 
edge modes propagating along and across the p-n interface.
In the bipolar regime the electron and hole modes 
can mix at the p-n boundary, leading to current partition
and quantized shot noise plateaus similar 
to those of conductance, while in the unipolar regime
transport is noiseless. These quantum 
Hall phenomena reflect the massless Dirac
character of charge carriers in graphene, 
with particle-hole interplay
manifest in mode mixing
and noise in the bipolar regime. 
\end{abstract}

\maketitle

The transport properties of graphene, 
2-dimensional sheets of graphite\,\cite{Geim07},
in particular the high carrier mobility
and tunability of transport characteristics,
make this material attractive for applications in 
nanoelectronics\,\cite{Son06,Rycerz07}.  
Various methods have been developed for patterning graphene 
sheets into prototype devices 
such as quantum dot transistors\,\cite{Geim07} 
and nanoribbons\,\cite{Chen07,Han07}, 
followed by demonstration of local control of carrier density
in a graphene sheet\,\cite{Huard07}.
Besides possible device applications, 
graphene junctions are predicted to host
new and exciting phenomena reflecting massless Dirac
character of carriers in this material, 
such as Klein tunneling\,\cite{Katsnelson06}, 
particle collimation\,\cite{Cheianov06}, 
quasibound states\,\cite{Silvestrov07}, 
and Veselago lensing\,\cite{Cheianov07}. 
In addition, interesting phenomena are expected in gated graphene bilayers, 
where the field-effect transport can be induced 
by tuning the gap at the Dirac point\,\cite{Ohta06}. 
These applications make gating of graphene a topic of great interest.

Recently, a graphene p-n junction with individual control of carrier density in
two adjacent regions with a pair of gates above and below it
was reported\,\cite{Marcus}. 
The density in each region could be varied across the neutrality point,
allowing pn, pp and nn
junctions to be formed at the interface.
The interface width was quite small owing to
$30\,{\rm nm}$ distance to the top gate and its sharp edge. 
Transport measurements, carried out in the quantized Hall
effect (QHE) regime at fields $3\,{\rm T}<B<8\,{\rm T}$, revealed ohmic 
two-terminal conductance
taking quantized values $g=6,2,3/2,1$ in the units of $e^2/h$. 
Along with the QHE plateaus with $g=2$ and $6$ in the unipolar regime,
quantized plateaus $g=1$ and $3/2$ of similar quality 
were observed in the bipolar regime.
While conductance of $6\frac{e^2}{h}$ and $2\frac{e^2}{h}$ 
is a hallmark of the integer QHE
in graphene\,\cite{Novoselov05,Zhang05},  
quantized conductance values $3/2$ and $1$ 
are unusual and call for explanation.

\begin{figure}
\includegraphics[width=3.5in]{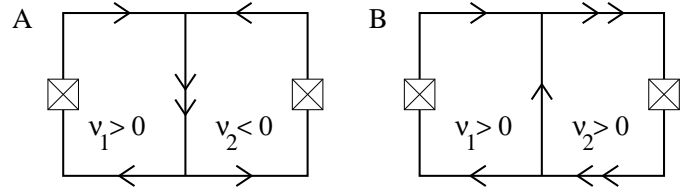}
\vspace{0.15cm}
\caption[]{ 
Schematic of QHE edge states for ({\textbf A}) bipolar regime,
and (\textbf{B}) unipolar regime of graphene junction. 
In case (\textbf{A}) the edge states counter-circulate in the n and p regions, 
bringing to the pn interface electrons and holes from different reservoirs. 
Mode mixing at the interface leads to the two-terminal conductance (\ref{eq:g=mn/(m+n)}).
In case (\textbf{B}), since the edge states circulate in the same direction
without backscattering or mixing, conductance is determined by the
modes permeating the whole system, $g={\rm min}(|\nu_1|,|\nu_2|)$.
}
\label{fig1}
\end{figure}

We interpret these observations
by linking them to the properties of the Dirac-like
carriers, which give rise to bipolar, 
electron and hole, QHE edge modes 
at the pn interface (Fig.\ref{fig1}).
The behavior at the interface is explained by
employing ideas from the theory of quantum-chaotic 
transport\,\cite{BeenakkerRMP,Baranger94,Blanter2000,Agam2000,Oberholzer01,Savin06}.
Although in our case the edge modes carry charge
along the pn interface
all in the same direction, in a chiral rather than
chaotic fashion, we argue that inter-mode scattering
within the pn interface region
gives rise to dynamics with features analogous to those
known for quantum-chaotic systems. 

In this analogy the QHE states at the
sample boundary play the role of perfect
lead channels of chaotic quantum dots\,\cite{BeenakkerRMP,Baranger94}, 
bringing charge to the pn interface and carrying it away into reservoirs.
However, several physical effects causing conductance fluctuations in chaotic
dots are absent in our case, leading to quantization 
of two-terminal conductance not known for the dots.
In particular, the effective lead channels are quantized 
more perfectly than in the dots, 
owing to backscattering suppression in QHE transport.
In addition, the quantum-mechanical interference effects 
which lead to sample-specific conductance fluctuations,
can be suppressed in our case due to self-averaging, 
as well as
dephasing and 
electron-electron scattering. 
Other effects that can affect the edge state transport 
at the pn interface are inter-mode
relaxation and coupling   
to electronic states in QHE bulk, causing dephasing in a manner similar
to the voltage probe model\,\cite{Buttiker_Vprobes}.
While these regimes yield similar results
for conductance, they will
manifest themselves differently in other characteristics, in particular
in electron shot noise\,\cite{Buttiker90}, which can 
be used for detailed characterization of transport mechanisms. 

\begin{figure}
\includegraphics[width=3.5in]{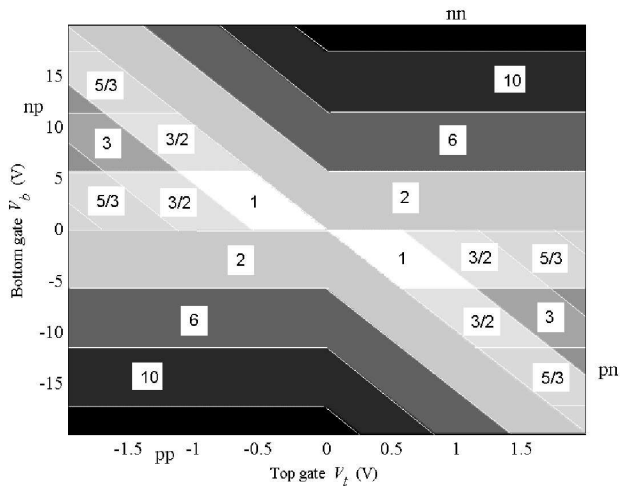}
\vspace{0.15cm}
\caption[]{ 
Two-terminal conductance 
{\em vs.} gate voltage, given by 
Eq.\,\ref{eq:g=mn/(m+n)}
in the bipolar case $\nu_1>0$, $\nu_2<0$
, and by
Eq.\,\ref{eq:g=min}
in the unipolar case ($\nu_{1,2}$ of equal sign).
The boundaries of QHE regions are specified by $\nu_{1,2}=0,\pm4,\pm8...$,
with the gate voltage dependence of $\nu_{1,2}$ 
given by Eq.\,\ref{eq:nu_12}.
Parameters used: distances to the top and back gates
$h=30\,{\rm nm}$, $d=300\,{\rm nm}$, 
magnetic length $\ell_B=10\,{\rm nm}$,
dielectric constant $\kappa=3$.
}
\label{fig2}
\end{figure}

Due to particle-hole symmetry of carriers in graphene, 
the QHE in this material occurs
symmetrically about the neutrality point 
at the densities $\nu= \pm 2,\pm 6,\pm 10...$\,\cite{Novoselov05,Zhang05}.
In each of these quantized states there are $n=|\nu|$ edge modes
propagating in different directions 
at $\nu>0$ and $\nu<0$\,\cite{Peres06,Abanin06}.
For the bipolar case, assuming QHE 
at densities $\nu_1>0$ and $\nu_2<0$ on either side of the boundary, 
this gives $|\nu_1|$ and $|\nu_2|$
edge modes circulating in opposite directions 
that 
merge to form a multi-mode edge state at the pn interface
(Fig.\ref{fig1}A). 
These modes supply to the pn interface 
particles from both the n and p reservoirs.
After propagating together along the interface these particles 
arrive at the sample boundary where they
are ejected into the edge modes which split up
and return to reservoirs.

The observed conductance quantization 
can be readily explained by assuming full mixing 
of these modes at the pn interface, such that
for each particle the probability to be ejected into any of 
the $N=|\nu_1|+|\nu_2|$ modes equals $p_N=1/N$ irrespective 
of its origin. The two-terminal conductance is then obtained
by multiplying $p_N$ by the numbers of the modes,
giving
\be\label{eq:g=mn/(m+n)}
g_{\rm pn}=\frac{|\nu_1||\nu_2|}{|\nu_1|+|\nu_2|}=1,\,\frac32,\,3,\,\frac53...
,
\ee
where $\nu_{1,2}=\pm2,\pm6,\pm10...$. This agrees
with the observed quantized values\,\cite{Marcus} (see Fig.\ref{fig2}).

\begin{figure}
\includegraphics[width=3.5in]{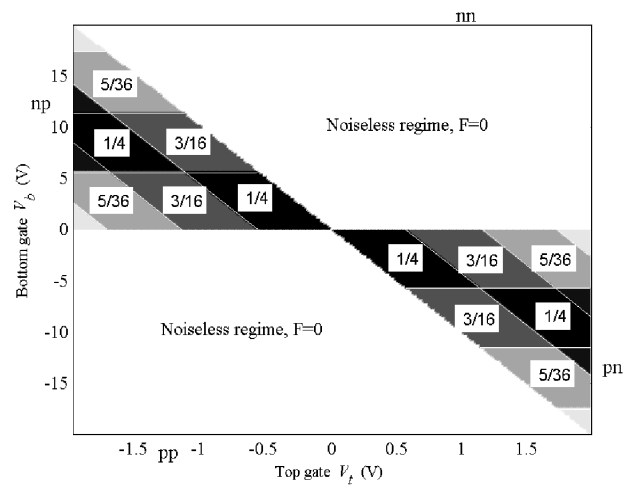}
\vspace{0.15cm}
\caption[]{ 
Shot noise Fano factor, Eq.\,\ref{eq:F_model1}, 
plotted {\em vs.} gate voltages for the same parameter values
as in Fig.\ref{fig2}. Noise is zero in the unipolar regime (pp or nn)
due the absence of current partition at the junction interface,
but finite in the bipolar regime
due to edge mode mixing at the pn interface.
}
\label{fig3}
\end{figure}

The character of QHE edge transport in the unipolar regime is 
quite different.
In this case, nn or pp, 
the edge modes in both regions circulate in the same direction. 
As a result, some modes are coupled to both reservoirs, 
while the others are connected to only one of the reservoirs 
(Fig.\ref{fig1}B). 
With backscattering suppressed by QHE, the conductance across 
the boundary is solely due to those edge modes 
that permeate the entire system,  
making contact with both reservoirs. 
This gives the observed nonclassical conductance
values
\be\label{eq:g=min}
g_{\rm nn}=g_{\rm pp}={\rm min}(|\nu_1|,|\nu_2|)=2,6,10...
,
\ee
$\nu_{1,2}=\pm2,\pm6,\pm10...$,
in agreement with
the known results for quantized conductance
of constrictions between different 
QHE states\,\cite{Haug88,Washburn88}.
The nondissipative character 
of transport in the unipolar regime, Eq.\,\ref{eq:g=min},
resulting from suppressed backscattering,
can be revealed by measuring noise. 
In the absence of current partitioning
inside the sample we expect only thermal Johnson-Nyquist noise
$S=2gk_{\rm B}T$ in this regime
but no shot noise contribution (see Fig.\ref{fig3}).

The conductance values given by Eqns.\,\ref{eq:g=min} and \ref{eq:g=mn/(m+n)} 
occur 
in a particular pattern\,\cite{Marcus} that can be described as follows
(see Fig.\ref{fig2}). 
Electron density in graphene induced by the back gate is 
$n_1=(\kappa/4\pi e)V_{\rm b}/d$, where $d$ is the distance to the gate,
$V_{\rm b}$ is voltage on it,
and $\kappa$ is dielectric constant. Similarly, 
in the locally gated region
we have $n_2=(\kappa/4\pi e)(V_{\rm b}/d+V_{\rm t}/h)$,
where $h$ and $V_{\rm t}$ are the distance to the top gate and voltage on it.
For the Landau level
filling factors $\nu_{1,2}=(hc/eB)n_{1,2}$ we find 
\be\label{eq:nu_12}
\nu_1=(\ell_B^2\kappa/2e)V_{\rm b}/d
,\quad
\nu_2=(\ell_B^2\kappa/2e)(V_{\rm b}/d+V_{\rm t}/h)
,
\ee
with $\ell_B=\sqrt{\hbar c/eB}$ the magnetic length.
The values $V_{\rm b}$, $V_{\rm t}$ corresponding to integer
QHE states are inside parallelograms with the boundaries approximately given by
$\nu_{1,2} = 0,\pm4,\pm8...$, 
as appropriate for the four-fold degenerate graphene
Landau levels\,\cite{Novoselov05,Zhang05}.
The resulting conductance pattern,
shown in Fig.\ref{fig2} for realistic parameter values,
strikingly resembles the experimental results\,\cite{Marcus}.


How is the conductance in Eq.\,\ref{eq:g=mn/(m+n)} 
affected by quantum-mechanical interference effects? 
Random matrix theory (RMT) 
of chaotic transport predicts \cite{BeenakkerRMP,Baranger94} ensemble-averaged conductance $\bar g=n_1n_2/(n_1+n_2+1-2/\beta)$,
where $n_{1,2}$ is the open channel number, and $\beta=1,2,4$ for the
three random matrix universality classes.
In our QHE case, with the channel numbers
$n_{1,2}=|\nu_{1,2}|$ and $\beta=2$, RMT predicts
$\bar g$ identical to Eq.\,\ref{eq:g=mn/(m+n)}.
Similarly, semiclassical description of transport    
in chaotic cavities\,\cite{Blanter2000}, where mixing is due to the 
dynamics in the cavity, yields conductance 
values close to the classical result for two conductors 
connected in series.

To clarify the origin of the mode mixing 
at the pn interface we studied electron density distribution
for the gate geometry used in Ref.\,\cite{Marcus}.
Numerical solution of Laplace problem for the electrostatic potential
in between the gates revealed that the pn density step is about $40\,{\rm nm}$
wide, a few times larger than the magnetic length at $B=8\,{\rm T}$.
Comparison to the known results\,\cite{Khaetskii94} 
for a compressible region sandwiched
between incompressible regions then suggests the presence at the pn interface
of additional QHE modes that, in the presence of disorder,
can facilitate inter-channel scattering and mixing.


In the fully coherent regime 
conductance would exhibit universal 
fluctuations, UCF.
The magnitude of UCF predicted for chaotic transport
(see Ref.\cite{Savin06})
in our case depends on the channel numbers as follows:
\be\label{eq:var(g)_model1}
{\rm var}(g)=\frac{\nu_1^2\nu_2^2}{(|\nu_1|+|\nu_2|)^2((|\nu_1|+|\nu_2|)^2-1)}
.
\ee
Applied to the observed plateaus with 
$(\nu_1,\nu_2)=(2,-2), (2,-6), (6,-2)$, Eq.\,\ref{eq:var(g)_model1}
indicates that 
these plateaus would not have been discernible in a system with fully developed
UCF. We therefore conclude that the observed quantization of $g$
depends on some mechanism that suppresses UCF.
For example, the suppression could easily be understood
if Thouless energy for the states at the pn interface was  
small compared to $k_{\rm B}T$. 
The reduced UCF would then result from averaging 
over
the $k_{\rm B}T$ energy interval. However,
the plateaus in \cite{Marcus} remain unchanged
when temperature is reduced 
from 4K to 250 mK,
making such a scenario unlikely.

The UCF suppression may signal a fundamental
departure of chiral QHE dynamics from that of the earlier studied systems. 
However, at this point we cannot exclude other, more mundane
explanations. In particular, time-dependent
fluctuations of system parameters can
supercede mesoscopic fluctuations, turning the observed time-averaged 
quantities into ensemble-averaged quantities.
This self-averaging could arise naturally due to
fluctuating electric field at the pn interface
induced by voltage noise on the gates. 
Another, more interesting explanation could be that UCF suppression 
indicates presence of dephasing due to the coupling of the chiral modes 
to the localized states in the bulk, or some other intrinsic mechanism.

Current partition due to mode mixing at the pn interface
will manifest itself in the finite shot noise intensity.
To evaluate noise, we note that mixing
of the reservoir distributions, no matter of what origin,
results in particle energy distribution of the form
\be
n(\epsilon)=\frac{|\nu_1|}{N}n_1(\epsilon)+\frac{|\nu_2|}{N}n_2(\epsilon)
\ee
which at small $k_{\rm B}T$ is a double step. 
In analogy with diffusive systems\,\cite{Nagaev92},
and chaotic cavities\,\cite{Blanter2000,Oberholzer01},
this distribution 
serves as a Kogan-Shulman-like extraneous source of current fluctuations,
\be\label{eq:n(1-n)}
J=\int n(\epsilon)(1-n(\epsilon))d\epsilon = \frac{|\nu_1||\nu_2|}{N^2}|V_{sd}|
.
\ee
We relate the noise source $J$ to the fluctuations 
of the two-terminal current by noting that,
since fluctuating current of intensity
$J$ is injected into each open channel, the current fluctuations 
flowing into the n and p regions will be $J_1=|\nu_1|J$
and $J_2=|\nu_2|J$.
Converting these fluctuations into voltage fluctuations
and adding the contributions of the n and p regions,
we find the voltage fluctuations induced between the reservoirs:
\be
\la \delta V^2\ra=\frac{J_1}{|\nu_1|^2}+\frac{J_2}{|\nu_2|^2}
=\lp \frac1{|\nu_1|}+\frac1{|\nu_2|}\rp J = \frac{|V_{sd}|}{N}
.
\ee
Current noise can now be obtained as $S=g^2\delta V^2$, 
where $g$ is the 
conductance (\ref{eq:g=mn/(m+n)}). 
It is convenient to characterize noise 
by the Fano factor $F=S/I$, describing noise suppression relative to
Poisson noise. We find
\be\label{eq:F_model1} 
F=\frac{|\nu_1||\nu_2|}{(|\nu_1|+|\nu_2|)^2}
=\frac14,\frac3{16},\frac5{36}...
,
\ee
where $\nu_{1,2}=2,6,10...$.
The result (\ref{eq:F_model1}) is identical in form to the shot noise 
Fano factors of chaotic cavities\,\cite{Blanter2000,Oberholzer01}. 
The Fano factor values (\ref{eq:F_model1})
should be contrasted with $F\approx 0.29$ predicted for a p-n junction
in the absence of 
magnetic field\,\cite{Cheianov06}.  

Another regime for noise is possible if electrons, 
while traveling along the 
pn interface, have enough time to 
transfer energy to each other via inelastic processes. 
This will occur if $\tau_{el}\ll L/v$,
where $\tau_{el}$ is the characteristic electron energy relaxation time,
$v$ is drift velocity and $L$ is the pn interface length. 
(A similar regime was analyzed
for diffusive\cite{Nagaev92} and chaotic\cite{Oberholzer01} transport.)
In this case,
the electron energy distribution is characterized 
by an effective temperature $T_{\rm eff}$ which 
is determined by the balance of 
the energy supplied from reservoirs
and electron thermal energy flowing out:
\be
\frac12\frac{|\nu_1||\nu_2|}{|\nu_1|+|\nu_2|}V_{sd}^2
=
\zeta(2)(|\nu_1|+|\nu_2|)k_{\rm B}^2T_{\rm eff}^2
,
\ee
$\zeta(2)=\pi^2/6$.
The extraneous fluctuations, Eq.\ref{eq:n(1-n)}, 
evaluated for the Fermi distribution with $T=T_{\rm eff}$, 
give $J=k_{\rm B}T_{\rm eff}$.
Repeating the reasoning that has led to Eq.\,\ref{eq:F_model1}
we find the noise intensity $S=gk_{\rm B}T_{\rm eff}$. This expression
resembles the Nyquist formula, except for the factor
of two missing because
the fluctuations (\ref{eq:n(1-n)}) occur only in the pn region but not 
in the leads. Since $T_{\rm eff}\propto V_{sd}$, 
this noise is linear in $V_{sd}$. Similar to
the $T=0$ shot noise, we characterize it by 
Fano factor $\tilde F=(3F)^{1/2}/\pi$,
with $F$ given by Eq.\ref{eq:F_model1}.

We finally note that noise can be used to test which of the UCF suppression mechanisms
discussed above, self-averaging or dephasing,
occur in experiment\,\cite{Marcus}. 
For coherent
transport noise exhibits mesoscopic fluctuations similar to UCF
which can be analyzed within RMT framework. 
In the absence of time reversal symmetry,
RMT
yields ensemble-averaged Fano factor
\be\label{eq:F_model1_exact} 
\bar F=\frac{|\nu_1||\nu_2|}{(|\nu_1|+|\nu_2|+1)(|\nu_1|+|\nu_2|-1)}
\ee
(see Eq.\,11 in Ref.\,\cite{Savin06}).
For $\nu_{1,2}=2,4,6...$
this gives $\bar F=4/15,12/63,36/143...$.
These values, expected when transport 
is coherent but self-averaged, are 
different from Eq.\ref{eq:F_model1}
obtained from incoherent mixing model. 

The quantized transport observed in graphene pn 
junctions\,\cite{Marcus} 
is of different character in the unipolar
and bipolar regimes. In the first case, transport is dissipationless
with conductance quantized to an integer. In the second case,
mode mixing at the pn interface creates a situation similar
to that studied in the quantum-chaotic
transport. Conductance quantized to fractional values
observed in Ref.\cite{Marcus} then
results from intrinsic or extrinsic suppression of UCF. 
These transport regimes can be unraveled using electron shot noise,
predicted to be finite in the
bipolar regime and zero in the unipolar regime,
with quantized plateau structure similar to that of conductance. 

We benefited from discussions with 
C. M. Marcus, L. Di Carlo, J. R. Williams, P. Kim and P. Jarillo-Herrero,
and from comments made by
Y. M. Blanter and E. V. Sukhorukov.
This work is supported by NSF MRSEC (DMR 02132802) and
NSF-NIRT DMR-0304019.

\end{document}